\documentclass[12pt,preprint]{aastex}

\usepackage{graphicx}
\usepackage{lscape}

\newcommand{\beq}{\begin{equation}}
\newcommand{\eeq}{\end{equation}}
\newcommand{\beqa}{\begin{eqnarray}}
\newcommand{\eeqa}{\end{eqnarray}}

\def\la{\lower.5ex\hbox{$\; \buildrel < \over \sim \;$}}
\def\ga{\lower.5ex\hbox{$\; \buildrel > \over \sim \;$}}

\begin{document}

\title{A Primeval Magellanic Stream and Others}

\author{P.~J.~E. Peebles}  
\affil{Joseph Henry Laboratories, Princeton University, Princeton, NJ 08544}
\author{R. Brent Tully}
\affil{Institute for Astronomy, University of Hawaii, 2680 Woodlawn Drive, Honolulu, H\,I 96822}

\begin{abstract}
The Magellanic Stream might have grown out of tidal interactions at high redshift, when the young galaxies were close together, rather than from later interactions among the Magellanic Clouds and  Milky Way. This is illustrated in solutions for the orbits of Local Group galaxies under the cosmological  condition of growing peculiar velocities at high redshift. Massless test particles initially near and moving with the Large Magellanic Cloud in these solutions end up with distributions in angular position and redshift  similar to the Magellanic Stream, though with the usual overly prominent leading component  that the Milky Way corona might have suppressed. Another possible example of the effect of conditions at high redshift is a model primeval stream around the Local Group galaxy NGC\,6822. Depending on the solution for Local Group dynamics this primeval stream can end up with position angle similar to the H\,I  around this galaxy, and a redshift gradient in the observed direction. The gradient is much smaller than observed, but might have been increased by dissipative contraction. Presented also is an even more speculative illustration  of the possible effect of initial conditions, primeval stellar streams around M\,31.\medskip

\end{abstract}

\maketitle

\section{Introduction}\label{sec:1}

The examples presented here of how intergalactic tidal streams could have been triggered by interactions among the young galaxies at high redshift, when they were all close together, are based on  solutions to a dynamical model for the Local Group (LG) under the condition of growing peculiar velocities at high redshift, in analogy to the growing mode of departure from a homogeneous  expanding universe in perturbation theory. In these solutions a reasonable approximation to the Magellanic Stream (MS) grew largely from the interaction of the young Large Magellanic Cloud (LMC) with its nearest massive neighbor at high redshift, the Milky Way (MW). That is, we arrive back at the picture introduced by Fujimoto  \& Sofue (1976) and Lin \& Lynden-Bell (1977), but applied at high redshift under cosmological initial conditions. 

The MS certainly is expected to have been affected by  subsequent tidal interactions, perhaps between the LMC and Small Magellanic Cloud (SMC), as noted by Fujimoto  \& Sofue (1976) and Lin \& Lynden-Bell (1977), and by interaction with the MW mass and corona (Meurer, Bicknell, 
\& Gingold 1985; Moore \& Davis 1994). Analyses of these effects (Gardiner \& Noguchi 1996; Mastropietro  et al. 2005; Connors, Kawata, \& Gibson 2006; Diaz \& Bekki 2011, 2012; Besla, Kallivayalil, Hernquist, et al. 2010, 2012; and references therein) show that tidal and hydrodynamical interactions at modest redshifts can produce plausible approximations to MS without reference to conditions at high redshift. Exploration of a primeval origin nevertheless is called for. Gravitational interactions among the young galaxies certainly are real, as exemplified by the cosmic web (Bond, Kofman, \& Pogosyan 1996). Exploration of the consequences in observations of the galaxies near us requires a prediction of how the young galaxies were positioned at high redshift. We seem to have that now for the LMC in a dynamical LG model constrained by initial conditions from cosmology and by the now considerable number of measurements of nearby galaxy redshifts, distances and proper motions (Peebles 2010; Peebles \& Tully 2013, PT). This invites exploration of the effect of primeval tidal interactions on the distribution of matter around the young LMC. The result is a credible first approximation to MS.

The LG dynamical model and methods of its solution are outlined in Section~\ref{sec:2} and discussed in more detail in PT and references therein. The dynamical actors and observational constraints are the same as in PT except that the actors that are meant to represent the effect of external mass are allowed the freedom to adjust angular positions as well as distances to aid the fit to the LG parameter constraints. The primeval streams presented in Section~\ref{sec:3} show purely gravitational motion of massless test particles, in the tradition of Toomre \& Toomre (1972). This simplifies the computation of streams in  this  preliminary exploration that might motivate more complete analyses that take account of hydrodynamics and self-gravitation. Section~\ref{sec:3.1} shows the evolution of the model for a primeval Magellanic Stream. It produces a reasonable-looking fit to the MS H\,I angular and redshift distributions without special parameter adjustments. This result motivates the exploration in Section~\ref{sec:3.2} of primeval streams around NGC\,6822. The observed H\,I envelope around this LG dwarf almost certainly is gravitationally bound to the galaxy,  a very different situation from MS. The results suggest that the H\,I envelope might have grown by dissipative contraction of a primeval H\,I stream, though substantiating that idea would require a considerable parameter search. A crude estimate of the situation is offered in Section~\ref{sec:4}. Section~\ref{sec:3.3} shows an even more speculative example, the development of streams around M\,31 from its interactions with M\,33, NGC\,185, and NGC\,147 at high redshift.  In the LG model solutions none of these galaxies passed close to M\,31 at modest redshifts, but streams form. This certainly cannot make the case for a primeval origin of streams around M\,31, because there are many other neighbors that could have produced streams at more modest redshifts, but it offers the possibility of a primeval component. We summarize our assessment of the results from the primeval stream models in Section~\ref{sec:4}.

\section{Dynamical Model and Solutions}\label{sec:2}

The starting assumption for the dynamical LG model is that the mass now concentrated around a galaxy was at high redshift in a patch whose motion may be traced by the position of its effective center of mass. This of course allows the galaxy to grow by accretion, provided it is accretion within the patch traced by the effective center. The initial condition is that the peculiar velocities of the mass patches are small and growing at high redshift. The condition that the galaxies end up where they are observed --- or else how they are observed to be moving --- presents mixed boundary conditions that are fitted by relaxation of the orbits to a stationary point of the action (in the NAM method introduced in Peebles 1989 and made more efficient in Peebles,  Tully \& Shaya 2011). In NAM solutions the equivalent of the decaying mode in linear perturbation theory is suppressed but not eliminated, as illustrated in Figure~1 in PT. It shows that, for the model parameters used in PT and here, peculiar velocities in the solutions are growing at redshift $z\la 20$  in a reasonable approximation to the wanted growing mode, while earlier than that the decaying mode that inevitably appears in a numerical solution dominates and diverges as $a(t)\rightarrow 0$. The advantage over a numerical integration back in time from given present positions and velocities is that NAM shifts domination of the decaying mode to high redshift where it seems likely to be harmless.

The NAM solutions are based on the $\Lambda$CDM cosmology with Hubble and matter density parameters
\beq
H_o=70\hbox{ km s}^{-1}\hbox{ Mpc}^{-1}, \qquad \Omega_m=0.27, \label{eq:cospars}
\eeq
where $\Omega_m$ represents the sum of masses in baryons and dark matter, the mass in radiation is neglected, space sections are flat, and Einstein's $\Lambda$ is constant. The numerical solutions trace back in time by expansion factor $1+z = 10$ (to redshift $z=9$) in 500 time steps uniformly spaced in the expansion parameter a(t). Numerical accuracy is checked by numerical integration forward in time in 5000 steps uniformly spaced in a(t) from positions and velocities at $1+z = 10$ from the action solution. The present positions and velocities from this forward integration generally agree with the action solution to better than 0.1~kpc and 0.3~km s$^{-1}$, apart from some solutions for Leo\,1, whose close passage of MW produces differences as large as 3~kpc and 5~km s$^{-1}$.

Solutions are found starting from random trial orbits with random initial assignments of distances, redshifts and masses within the nominal uncertainties, the orbits relaxed to a stationary point of the action, and the parameters then iteratively adjusted and relaxed to a stationary point to improve the  fit to the measurements of LG redshifts, distances, luminosities, and peculiar velocities. The mixed boundary conditions allow many discretely different solutions; we  choose the more plausible ones by comparison to the data. More details are in PT.

Table~1 names the LG galaxies in the dynamical model. The adopted LG parameter values and their measured or estimated uncertainties are entered under the headers ``catalog'' (or ``cat''). Entered under the headers ``solution'' are the parameter values in three numerical solutions to the dynamical model, ordered by the goodness of fit to the data. The catalog distances and their uncertainties, redshifts, and luminosities are from the Local Universe (LU) catalog maintained and provided on-line by Tully.\footnote{http://edd.ifa.hawaii.edu select Local Universe (LU) catalog} The adopted nominal uncertainty in each redshift, 10 km~s$^{-1}$, is meant to allow for possible motion of the galaxy of stars relative to its dark matter halo. The nominal catalog masses (baryonic plus dark matter) are computed from the K-band luminosities using mass-to-light ratio $M/L_K=50M_\odot/L_\odot$, meaning the nominal value of $M/L_K$ in Table~1 is 50. The nominal uncertainties in the LG galaxy masses are placed on a logarithmic scale, with a factor of 1.5 at one standard deviation. The exception is MW, whose nominal mass ratio to M\,31 is unity with a factor of 1.1 at one standard deviation  (PT eqs.~[5] and~[6]).  The nominal rms galaxy peculiar velocity at $1+z=10$ is taken to be $v_i=50$~km~s$^{-1}$ for LG and external actors. This is roughly what might be expected from the growth to rms peculiar velocities several times that at the present epoch. The velocity of the Sun relative to the local standard of rest is from Sch{\"o}nrich, Binney \& Dehnen  (2010), with no allowance for uncertainty in this relatively small term. The circular velocity of the local standard of rest has catalog value $v_c=230\pm 10$~km~s$^{-1}$. The mass distribution in each actor is rigid and spherical with density run $\rho\propto r^{-2}$ cut off at the radius that produces the model mass for given $v_c$. Assigned circular velocities without uncertainties are $v_c=250$~km~s$^{-1}$ in M\,31 and $v_c=100$~km~s$^{-1}$ in all the other actors except MW.

\begin{landscape}
\begin{table}[htpb]
\centering
\begin{tabular}{rccccrrrrrrrrrrrrr}
\multicolumn{18}{c}{Table 1: Local Group Model Distances, Redshifts, Masses, and Initial Velocities}\\
\noalign{\medskip}
\tableline\tableline\noalign{\smallskip}
 Name  & \multicolumn{4}{c}{Distance} &&
      \multicolumn{4}{c}{redshift} &&
      \multicolumn{3}{c}{$M/L_K$} && \multicolumn{3}{c}{$v_i$} \\
      \noalign{\smallskip}
      \cline{2 - 5}
       \cline{7 - 10}
       \cline{12 - 14}
        \cline{16 - 18}
      & catalog & \multicolumn{3}{c}{solution} &&  cat &  \multicolumn{3}{c}{solution}&& 
        \multicolumn{3}{c}{solution} &&  \multicolumn{3}{c}{solution} \\
      \cline{3 - 5} \cline{8 - 10} \cline{12 - 14} \cline{16 - 18}
      & & 1 & 2 & 3 &&& 1 & 2 & 3 && 1 & 2 & 3  && 1 & 2 & 3 \\
   \tableline
\noalign{\smallskip}
  MW      & $   0.0085 \pm   0.0001 $ & $   0.008$ & $   0.008 $ & $   0.008$ && $  -11$ & $  -11$ & $  -11$ & $  -11$ &&    33 &    36 &    40 &&    65 &    62 &    60 \\
 M31     & $   0.770 \pm   0.040 $ & $   0.747$ & $   0.779 $ & $   0.829$ && $ -301$ & $ -286$ & $ -294$ & $ -291$ &&    25 &    27 &    31 &&    46 &    32 &    32 \\
 LMC     & $   0.049 \pm   0.006 $ & $   0.060$ & $ {\it 0.062} $ & ${\it 0.064}$ && $  271$ & $  266$ & $  271$ & $  269$ &&    34 &    45 &    40 &&    64 &    59 &    59 \\
 M33     & $   0.910 \pm   0.050 $ & $   0.905$ & $   0.815 $ & $   0.851$ && $ -180$ & $ -197$ & $ -192$ & $ -187$ &&    47 &    95 &    60 &&    86 &    76 &    69 \\
 I10     & $   0.790 \pm   0.040 $ & $   0.830$ & $   0.827 $ & $   0.789$ && $ -348$ & $ -350$ & $ -348$ & $ -347$ &&    50 &    44 &    41 &&    43 &    39 &    61 \\
 N185    & $   0.640 \pm   0.030 $ & $   0.639$ & $   0.639 $ & ${\it 0.559}$ && $ -227$ & $ -220$ & $ -226$ & $ -207$ &&    49 &    51 &    53 &&    52 &    47 &    52 \\
 N147    & $   0.730 \pm   0.040 $ & $   0.753$ & $   0.781 $ & $   0.779$ && $ -193$ & $ -191$ & $ -190$ & $ -188$ &&    48 &    49 &    43 &&    41 &    41 &    90 \\
 N6822   & $   0.510 \pm   0.030 $ & $   0.509$ & $   0.511 $ & $   0.505$ && $  -57$ & $  -56$ & $  -60$ & $  -60$ &&    49 &    49 &    50 &&    76 &    81 &    73 \\
 LeoI    & $   0.260 \pm   0.010 $ & $   0.260$ & $   0.262 $ & $   0.266$ && $  284$ & $  284$ & $  273$ & $  284$ &&    49 &    50 &    49 &&    57 &    92 &    98 \\
 LeoT    & $   0.410 \pm   0.020 $ & $   0.401$ & $   0.406 $ & $   0.412$ && $   35$ & $   42$ & $   37$ & $   33$ &&    49 &    50 &    49 &&    73 &    79 &    88 \\
 Phx     & $   0.410 \pm   0.020 $ & $   0.419$ & $   0.420 $ & $   0.415$ && $  -13$ & $  -22$ & $  -23$ & $  -17$ &&    49 &    49 &    49 &&    62 &    64 &    60 \\
 LGS3    & $   0.650 \pm   0.130 $ & $   0.705$ & $   0.849 $ & $   0.712$ && $ -281$ & $ -281$ & $ -277$ & $ -281$ &&    49 &    49 &    50 &&    55 &    47 &    53 \\
 CetdSp  & $   0.730 \pm   0.040 $ & $   0.728$ & $   0.734 $ & $   0.730$ && $  -87$ & $  -85$ & $  -89$ & $  -87$ &&    50 &    49 &    50 &&    66 &    67 &    64 \\
 LeoA    & $   0.740 \pm   0.110 $ & $   0.572$ & $   0.575 $ & $   0.545$ && $   28$ & $   34$ & $   33$ & $   33$ &&    49 &    50 &    50 &&    63 &    67 &    67 \\
 I1613   & $   0.750 \pm   0.040 $ & $   0.752$ & $   0.743 $ & $   0.751$ && $ -238$ & $ -239$ & $ -229$ & $ -239$ &&    50 &    50 &    50 &&    71 &    60 &    74 \\
 \noalign{\smallskip}
\tableline
\noalign{\smallskip}
\multicolumn{4}{l}{units: Mpc,  km s$^{-1}$,  $M_\odot/L_\odot$} \\
\end{tabular}
\end{table}

\end{landscape}
\begin{landscape}

\begin{table}[htpb]
\centering
\begin{tabular}{lrrrrrrrrr}
\multicolumn{10}{c}{Table 2: Proper Motions}\\
\noalign{\medskip}
\tableline\tableline\noalign{\smallskip}
Name  & \multicolumn{4}{c}{$\mu_\alpha$} && \multicolumn{4}{c}{$\mu_\delta$}\\
\cline{2 - 5} \cline{7 - 10}
 & \multicolumn{1}{c}{catalog}  & \multicolumn{3}{c}{solution} && \multicolumn{1}{c}{catalog} & \multicolumn{3}{c}{solution} \\
 \cline{3 - 5} \cline{8 - 10}
 && \multicolumn{1}{c}{1} & \multicolumn{1}{c}{2} & \multicolumn{1}{c}{3} 
 &&& \multicolumn{1}{c}{1} & \multicolumn{1}{c}{2} & \multicolumn{1}{c}{3} \\
\tableline
 \noalign{\smallskip}
  M31    & $   0.0440 \pm   0.0130 $ & $   0.0216$ & $   0.0246$ & $   0.0188$ && $  -0.0320 \pm   0.0120 $ & $  -0.0162$ & $  -0.0216$ & $  -0.0221$ \\
 LMC    & $   1.9100 \pm   0.0200 $ & $   1.9160$ & $   1.9155$ & $   1.9162$ && $   0.2290 \pm   0.0470 $ & $   0.2571$ & $   0.2513$ & $   0.2498$ \\
 M33    & $   0.0230 \pm   0.0060 $ & $   0.0116$ & $   0.0147$ & $   0.0201$ && $   0.0020 \pm   0.0070 $ & $   0.0092$ & ${\it 0.0170}$ & ${\it 0.0178}$ \\
 I10    & $  -0.0020 \pm   0.0080 $ & $ -{\it 0.0231}$ & $ -{\it 0.0218}$ & $ -{\it 0.0184}$ && $ 0.0200 \pm   0.0080 $ & $   0.0255$ & $   0.0298$ & ${\it 0.0408}$ \\
 LeoI   & $  -0.1140 \pm   0.0295 $ & $ -{\it 0.0478}$ & $  -0.1110$ & $  -0.1222$ && $  -0.1256 \pm   0.0293 $ & $  -0.1659$ & $  -0.1800$ & $  -0.1833$ \\
\tableline
 \noalign{\smallskip}
\multicolumn{3}{l}{unit: milli arc sec y$^{-1}$}\\
\end{tabular}
\end{table}

\begin{table}[htpb]
\hspace{-5mm}
\begin{tabular}{lrrrrrrrrrrrrrrrrrrrrrrr}
\multicolumn{24}{c}{Table 3: External Actors}\\
\noalign{\medskip}
\tableline\tableline\noalign{\smallskip}
& \multicolumn{8}{c}{supergalactic coordinates, SGL, SGB} && \multicolumn{4}{c}{distance} && \multicolumn{4}{c}{redshift} && \multicolumn{4}{c}{mass}\\
 \cline{2 - 9} \cline{11 - 14}  \cline{16 - 19} \cline{21 - 24}
& \multicolumn{2}{c}{catalog}  &  \multicolumn{6}{c}{solution}   && \multicolumn{1}{c}{cat}  &  \multicolumn{3}{c}{solution $\delta D$} 
 	&& \multicolumn{1}{c}{cat}  &  \multicolumn{3}{c}{solution}&& \multicolumn{1}{c}{cat}  &  \multicolumn{3}{c}{solution}   \\
  \cline{4 - 9} \cline{12 - 14} \cline{17 - 19}  \cline{22 - 24}
  &&& \multicolumn{2}{c}{1}  & \multicolumn{2}{c}{2}  & \multicolumn{2}{c}{3}  
   &&& \multicolumn{1}{c}{1}  &   \multicolumn{1}{c}{2}  & \multicolumn{1}{c}{3}
   &&& \multicolumn{1}{c}{1}  & \multicolumn{1}{c}{2}  & \multicolumn{1}{c}{3}  
   &&& \multicolumn{1}{c}{1}  &   \multicolumn{1}{c}{2}  & \multicolumn{1}{c}{3}\\
   \noalign{\smallskip}
\tableline
\noalign{\smallskip}
Scl &$  271$&$   -5$&$  269$&$    5$&$  268$&$   -3$&$  278$&$   -1$&&  3.6&  0.7&  0.3&  0.5&&$  242$&$  210$&$  220$&$  212$&&   51&   30&   26&   23 \\
Maf &$    1$&$    1$&$    3$&$    3$&$    3$&$    4$&$    1$&$    2$&&  3.6&  0.4&  0.3&  0.4&&$    9$&$   -9$&$   -4$&$   -4$&&   70&   27&   21&   29 \\
M81 &$   41$&$    1$&$   54$&$  -13$&$   60$&$  -11$&$   63$&$   -7$&&  3.7&{\it 1.8}&{\it 2.0}&{\it 2.0}&&$   46$&$   -1$&$   22$&$   -7$&&   71&  101&   71&   93 \\
Cen &$  161$&$   -4$&$  158$&$    0$&$  160$&$   -3$&$  161$&$   -3$&&  4.0&  0.3&  0.1&  0.1&&$  518$&$  510$&$  515$&$  527$&&  119&   51&   67&   51 \\
 \noalign{\smallskip}
\tableline
\noalign{\smallskip}
\multicolumn{8}{l}{units:  Mpc, km s$^{-1}$, $10^{11}M_\odot$} \\
\end{tabular}
\end{table}
\end{landscape}

Table 2 lists proper motions, where $\mu_\alpha$ is the motion in the direction of increasing right ascension and $\mu_\delta$ is the motion in the direction of increasing declination. The measurement and uncertainty for M31 is from Sohn, Anderson  \& van der Marel (2012), for LMC from Kallivayalil et al. (2013), for M33 from Brunthaler et al. (2005),  for IC\,10 from Brunthaler et al. (2007), and for LeoI from Sohn, Besla, van der Marel, et al. (2012).

The external actors named in Table 3 are meant to give a phenomenological description of the effect of the external mass distribution on LG by allowing their present positions and masses to float to aid the model fit to the catalog LG parameters. In a departure from PT, the angular positions as well as distances of these actors are allowed to float. The nominal  angular positions (columns~2 and~3 in Table~3) are luminosity-weighted means for the galaxies concentrated around the Sculptor group, the Maffei-IC\,342 system, the M\,81 group, and the Centaurus-M\,94 system. The positions are given in supergalactic coordinates, because the nearby galaxies outside LG are concentrated near the supergalactic plane. The distances $\delta D$ between the three-dimensional positions in the catalog and the model solutions have nominal allowed rms value 0.5\,Mpc. The nominal redshifts are luminosity-weighted means, with adopted uncertainties 50~km~s$^{-1}$. The catalog masses are computed from the sums of K-band luminosities with $M/L_K=50M_\odot/L_\odot$, and the mass uncertainties are on a logarithmic scale with one standard deviation at a factor of two difference between catalog and model solution. 

The measured or adopted uncertainties in the catalog parameters are treated as standard deviations in a $\chi^2$ sum of squares of differences between model and catalog values divided by standard deviations. There are 69 LG parameters: 14 distances,  14 redshifts, 15 masses, 10 components of proper motion, the MW circular velocity, and 15 primeval velocities. (The last are more properly counted as 45 primeval velocity components, each with a Gaussian velocity distribution, less three components because the center of mass is at rest, but this is too fine for the present purpose). There are 24 external actor parameters: the redshift, mass, primeval velocity, and three components of present position for each of the four  actors. Solutions 1 through 3 have $\chi^2=92$, 98, and 114, close to the total of 93 parameters in $\chi^2$. This is not very meaningful, however, for two reasons. First, many of the nominal standard deviations are at best only informed guesses. Second, the multiple solutions allowed by the mixed boundary conditions allow multiple choices among which we choose those with the smallest $\chi^2$. That is, if model, measurements and standard deviations were accurate enough for a meaningful value of $\chi^2$  we would expect it to be less than 93. The sums over LG parameters alone are $\chi^2_{\rm LG}=70$, 76, and 89. These are not much larger than the 69 LG parameters, but again one would have expected smaller because the solutions were chosen for their fit to the catalog parameters, and because the external parameters were adjusted to reduce $\chi^2_{\rm LG}$. If it were supposed that the external actor parameters are in effect free, because their constraints are quite loose, one might  expect $\chi^2_{\rm LG}=69 - 16 = 53$ (discounting only the masses and present positions, which most matter for LG orbits ), which would put the reduced $\chi^2_{\rm LG}$ values at about 1.5, well above statistical expectation.

The initial peculiar velocities of the external actors are less than about 25~km~s$^{-1}$, which seems acceptably small. The numerical solutions put the mass of the M81 actor at or above its nominal value and the other masses below nominal, and  M81 is placed at heliocentric distance $\sim 2$\,Mpc, well short of the LU distance to the galaxy M\,81, $3.65\pm 0.18$ Mpc, while the other three actors are placed close to their catalog positions, at median $\delta D\sim 0.3$~Mpc. Parameter adjustments are allowed to move the external actors away from the plane, but the model positions now and at $1+z=10$ are close to the plane, supergalactic latitude SGB close to zero (columns 4 to 9 in Table~3). A possibly significant exception is that the solutions prefer the M81 actor below the plane. This is in the direction that would help compensate for the striking scarcity of galaxies --- and likely mass --- in the Local Void immediately above the plane. We hope to investigate this and other aspects of the influence on LG of the external mass distribution in due course.

The parameters in the numerical solutions in Tables~1 to~3 that differ from catalog by more than two nominal standard deviations are entered in italics. (There are no 3-$\sigma$ differences in LG parameters.) The model prefers a long LMC distance, at $1.8\sigma$ in Solution~1, $2.5\sigma$ in Solution~3. Since measurements of this distance have been thoroughly examined we expect the discrepancy indicates a systematic error in the model, perhaps in the simple approximation to the mass distribution in MW. The other 2-$\sigma$ discrepancy in Table~1 is the short distance to NGC\,185 in Solution~3. None of the model solutions fit both catalog proper motion components of IC\,10 to better than two standard deviations, and only Solution~1 fits the proper motion of M\,33, but it fails the proper motion of Leo\,I at two standard deviations. Here again the problem certainly may be with the model, but since these proper motion measurements have not been so thoroughly reconsidered one may imagine some of the errors assigned to these difficult measurements are underestimates. 

In Solutions~1 to~3 respectively the MW circular velocity is $v_c = 229$, 238, and 246~km~s$^{-1}$. The increase with decreasing quality of fit to the measurements may be accidental. The preference for a value larger than the standard  220~km~s$^{-1}$ is in the direction of but smaller than that found by Reid, Menten, Zheng, Brunthaler, et al. (2009). The preference for greater mass in MW than M\,31 is present also in the larger number of solutions in PT with generally poorer fits to the constraints. In models~1 to~3 the MW masses in units of $10^{11}M_\odot$ are 15.2, 16.3, and 18.3, respectively, and the M\,31 masses are 12.9, 13.6 and 15.8. If the relative motion of MW and M\,31 were not affected by other actors the sum of the derived masses would decrease with increasing $v_c$ because the derived galactocentric redshift of M\,31 decreases with  increasing $v_c$. The trend the other way in the three solutions is not inconsistent with this argument because the other actors significantly affect the relative motion of MW and M\,31,  curving the orbit. 

In future work aimed at improving the LG dynamical model we expect to use more realistic descriptions of the mass distributions within MW and M\,31. It may help also to let the characteristic radius of the mass distribution within M\,31 be an adjustable parameter (in analogy to the adjustment of $v_c$ in a truncated limiting isothermal sphere). And, perhaps most important, the treatment of the effect of the external mass distribution on LG should more closely refer to the observed external galaxy distribution and peculiar motions.

The numerical solutions are not formally statistically consistent with the full catalog of parameters within their uncertainties. That is not surprising, because the dynamical model is crude and the catalog likely to contain errors. It is encouraging that the solutions match a considerable number and variety of constraints with relatively few discrepancies beyond two nominal standard deviations and, in LG, none beyond $3\sigma$. This degree of fit to the constraints argues that we have a reasonably secure basis for exploration of the effect of initial conditions on the development of streams within the Local Group. 

\begin{figure}[htpb]
\begin{center}
\includegraphics[angle=0,width=2.75in]{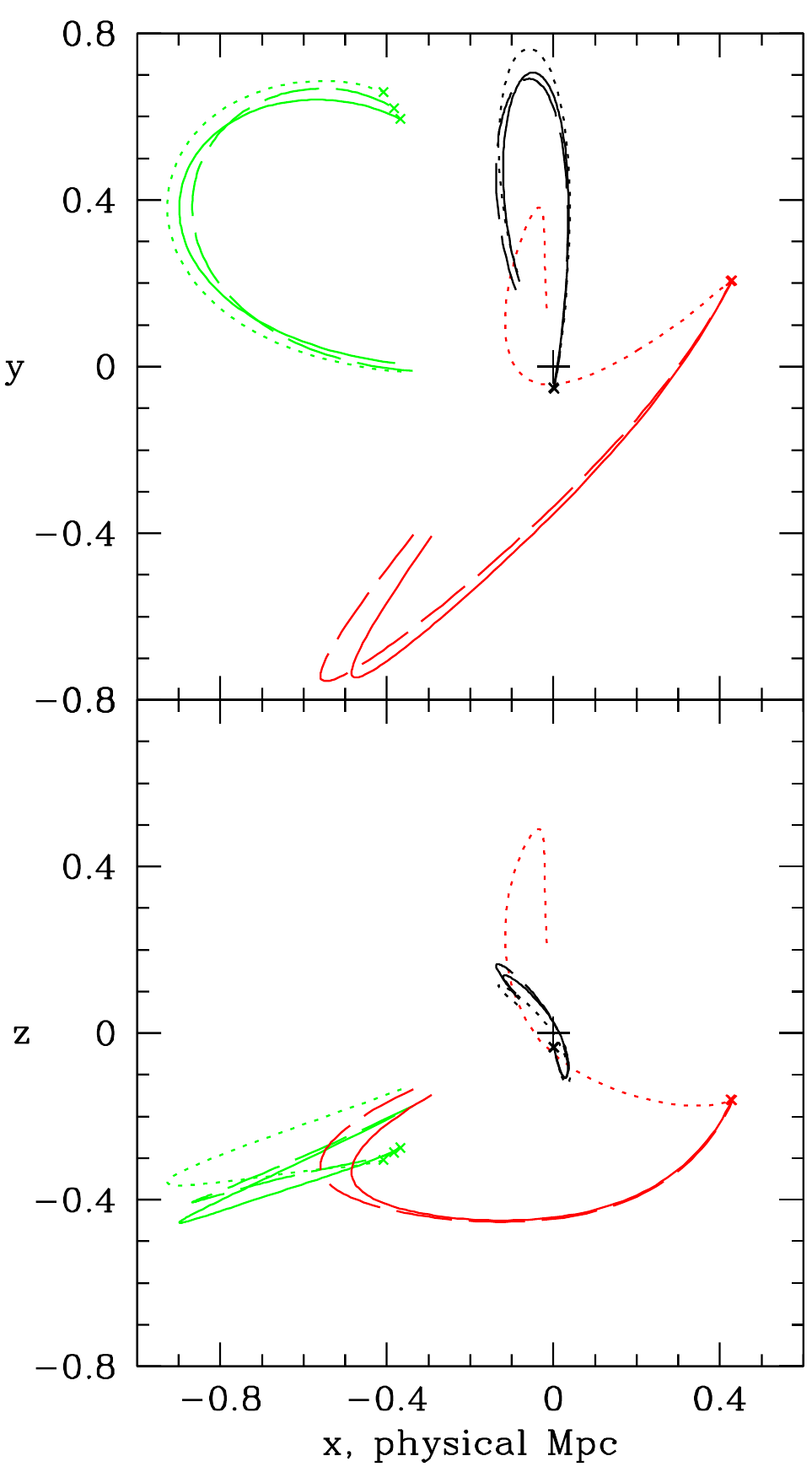} 
\caption{\label{Fig:1} \footnotesize  Orbits around MW, at the origin, for  LG galaxies LMC (black),  NGC\,6822 (red), and M\,31 (green). Solution 1 is plotted as the solid lines, 2 as long dashes, and 3 as dots.}
\end{center}
\end{figure}

\section{Primeval Streams}\label{sec:3}

Figure~\ref{Fig:1} shows model solutions for the motions of galaxies relative to MW. The right-hand coordinate system is  galactic, with the $z$-axis at $b=90^\circ$ and the $x$-axis at $b = l = 0$. The lengths are physical. Positions are plotted relative to MW at the plus sign. We consider possible examples of remnant primeval streams around LMC, plotted in black, and NGC\,6822, plotted in red. The green curves show the relative position of the other massive LG actor, M\,31. Present positions are at the crosses, and the other ends of the orbits show the young galaxies moving apart at redshift $z=9$. The solid lines show Solution~1 in Tables~1 to~3, the dashed lines Solution 2, and dotted, 3. These solutions are at different stationary points of the action and local minima of $\chi^2$. The three orbits of LMC relative to MW are quite similar, and they are similar too in the greater number of solutions  in PT. The apparently well-determined initial situation of LMC invites our exploration of the effect of the initial conditions on a cloud of test particles that might approximate the behavior of an H\,I envelope. The motion of NGC\,6822 relative to MW is similar in Solutions~1 and~2, quite different in~3.  This illustrates the multiple solutions allowed by the mixed boundary conditions. For this galaxy  an assessment of the situation has to guide selection of the more likely solution. 

The test particles in a model stream around a chosen LG actor move in the given gravitational field of the solution for the actors with mass, making it easy to accumulate a dense sample of test particle paths. The test particles are placed at $z=9$ uniformly at random within the gravitational radius $x_g$ of the actor, where 
\beq
x_g = (2GM/\Omega H_o^2)^{1/3}. \label{eq:gravradius}
\eeq
The comoving length $x_g$ is normalized to the physical radius at the present epoch. The galaxy mass, $M$, is the same as the mass within $x_g$ in a homogeneous mass distribution at the cosmic mean density. This means that a test particle closer than $x_g$ tends to be gravitationally attracted to the actor, in comoving coordinates, and at $x>x_g$ a test particle tends to be pulled away. The radii $x_g$ also are the characteristic separations of the actors at high redshift, where the peculiar accelerations of the actors are bounded by the condition that the orbits approximate the expansion of a near-homogeneous mass distribution. 

In a model stream the test particles are initially at rest in comoving coordinates relative to the chosen actor, meaning the particles initially are streaming away from the actor with the general expansion of the universe. The condition for capture of a test particle by the chosen actor varies with direction as well as distance relative to $x_g$, of course. Trials that take account of this by rejecting test particles with initial relative peculiar accelerations directed away from the chosen actor, and trials with initial velocities set to what is derived from the initial peculiar gravitational acceleration in linear perturbation theory, do not to produce very different streams, so for simplicity these refinements are not used in the results presented here.

\begin{figure}[htpb]
\begin{center}
\includegraphics[angle=0,width=5.25in]{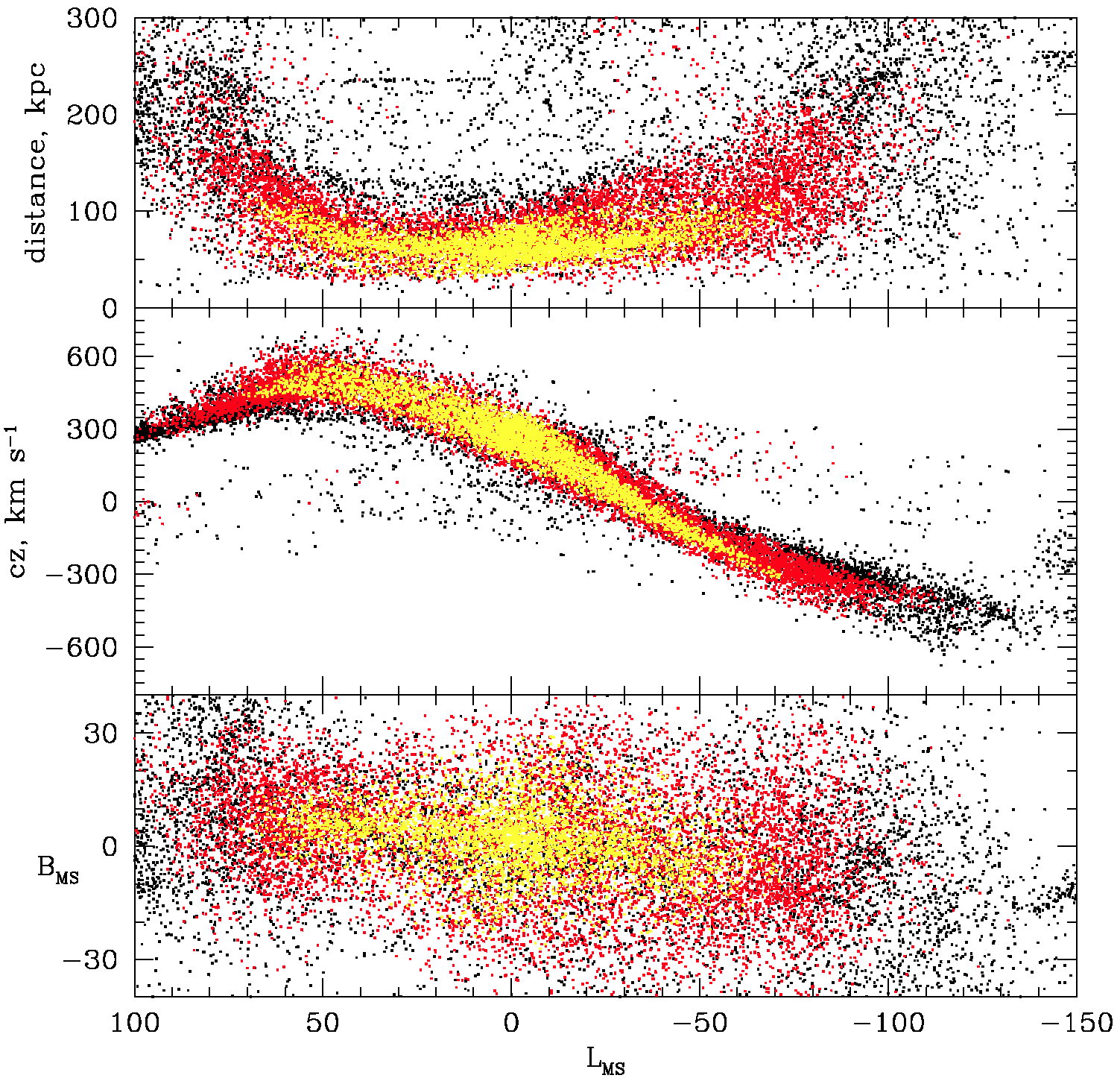} 
\caption{\label{Fig:2} \footnotesize  The Magellanic Stream in Solution~1. Particles initially less than 41~kpc from LMC  are plotted in yellow, those initially at $41 < r < 62$ kpc in red, and those at $62 < r < 82$ kpc in black.}
\end{center}
\end{figure}

\subsection{The Magellanic Stream}\label{sec:3.1}

Figure~\ref{Fig:2} shows present positions and redshifts of the test particles that are uniformly distributed within $r_g$ around LMC at $z=9$ in Solution~1.  The plots derived from the other two solutions look quite similar (and the gravitational radii defined in eq.~[\ref{eq:gravradius}] are similar; the physical values at $1+z=10$ are $r_g=82$, 90, and 87~kpc in Solutions~1 to~3). The sharp cutoff in the initial distribution of test particles can produce features in the present distribution of the particles plotted in black that are at initial physical distance $0.75\, r_g$ to $r_g$ from LMC, but the effect is not prominent here. There is not  much indication of orbit mixing; the appearance rather is that the initial distribution has been smeared by a smooth velocity field.

Figure~\ref{Fig:2} is plotted in the MS coordinates defined by Nidever, Majewski, \& Burton (2008), where LMC is at the origin, the stream is centered near $B_{\rm MS}=0$, and the trailing stream is at $L_{\rm MS}<0$. Figure~\ref{Fig:2} can be compared to Figure~8 in Nidever et al. (2010), which shows in these coordinates the measured H\,I angular distribution and the heliocentric redshift as a function of $L_{\rm MS}$. The orientation of the model stream and the variation of the heliocentric redshift with $L_{\rm MS}$ are close to what is observed, though the model lacks the fine structure in MS, including the SMC, and the leading stream is much too prominent. Figure~\ref{Fig:2} also is  similar to other MS models, including Figures~6 and~7 in Gardiner \& Noguchi (1996); Figures~9 and~12 in Mastropietro  et al. (2005); Figures~7 and~9 in Connors, Kawata, \& Gibson (2006); Figure~7 in Diaz \& Bekki (2012); and Figures~7 and~9 in Besla et al. (2012).  In Figure~\ref{Fig:2} the distance to MS is nearly constant from $-50^\circ \la L_{\rm MS}\la 50^\circ$ at about 60~kpc. The variation of distance with $L_{\rm MS}$ is similar in Connors, Kawata, \& Gibson, and in Diaz \& Bekki, while Besla et al. show a more rapid decrease of distance with increasing $L_{\rm MS}$. An observational check does not seem likely, however. The notable overall similarity of results in a variety of models is discussed in Section~\ref{sec:4}.

\begin{figure}[htpb]
\begin{center}
\includegraphics[angle=0,width=6.5in]{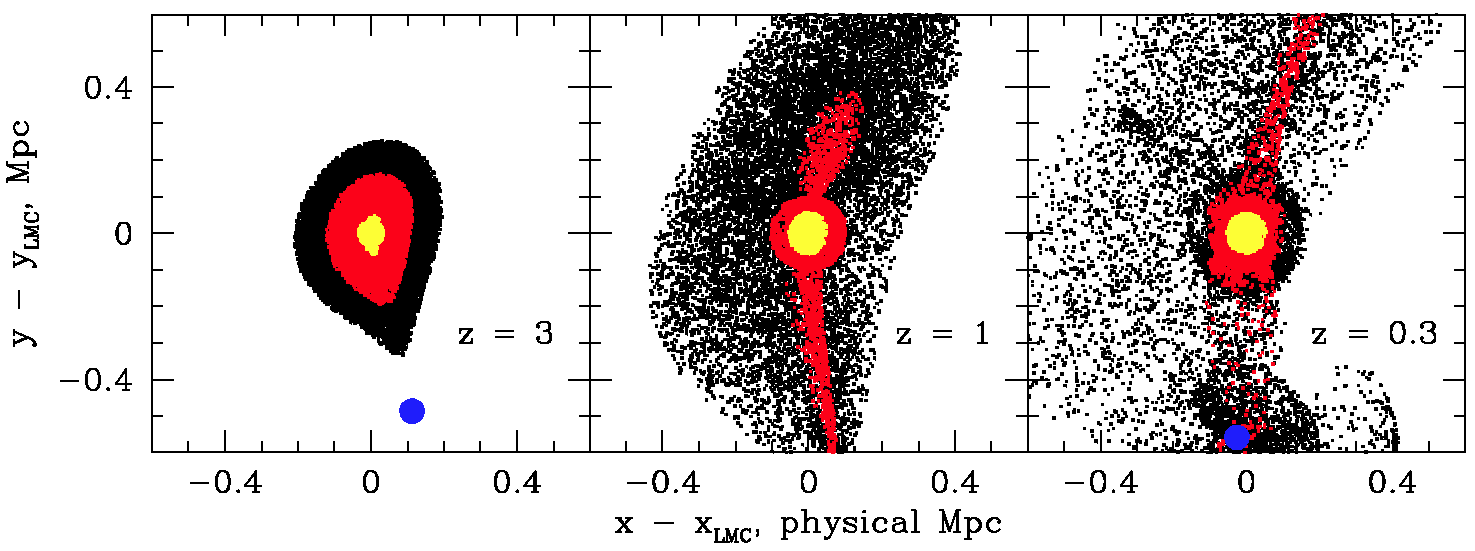} 
\caption{\label{Fig:3} \footnotesize  Evolution of the Magellanic Stream model in Figure~\ref{Fig:2}, with the same color scheme. MW is at the blue circle. Coordinates are galactic, lengths physical.}
\end{center}
\end{figure}

Figure~\ref{Fig:3} illustrates the evolution of the cloud of test particles around LMC in Solution 1.  Solutions~2 and~3 look similar.  The effect of the sharp cutoff in the distribution of black particles is more apparent here than in Figure~\ref{Fig:2}. At $z=9$ the physical distance between MW and LMC is 250~kpc, comparable to the MW gravitational radius 220~kpc (eq.~[\ref{eq:gravradius}]) at $z=9$, and M31 is about 500~kpc from LMC. At $z=3$, in the left-hand panel, MW is at the blue circle near the bottom edge. Its distance from LMC has about doubled, growing slightly less than the factor 2.5 general expansion. M\,31 is to the left, outside the boundary of the figure. The physical width of the distribution of the initially innermost particles plotted in yellow is about the same at $z=3$ as at $z=9$, while the outer envelope marked by the black particles has expanded by about as much as the general expansion. The elongated distribution of test particles at $z=3$ points to MW, even among the initially innermost  particles particles shown in yellow. 

In the central panel in Figure~\ref{Fig:3}, at $z=1$, MW is close to its maximum separation from LMC, 700~kpc. It is below the panel and positioned about in line with the prominent red stream and the long axis of the slightly eccentric distribution of yellow particles. MW reappears in the right-hand panel, at $z=0.3$. By this time some of the test particles are concentrated around MW. The band of black particles in the upper left part of the right-hand panel is suggestive of folding in singe-particle phase space. The yellow particles that initially were less than 41~kpc from LMC are still concentrated around LMC at about this radius at $z=0.3$. They end up smeared into the yellow stream in Figure~\ref{Fig:2}. 

The point of Figure~\ref{Fig:3} is that the cloud of test particles around LMC carries some memory of its interaction with MW when they were close, at high redshift. This is seen in features in the three panels of Figure~\ref{Fig:3} and, at $z=0$, in the stream in Figure~\ref{Fig:2}. 

\begin{figure}[htpb]
\begin{center}
\includegraphics[angle=0,width=4.in]{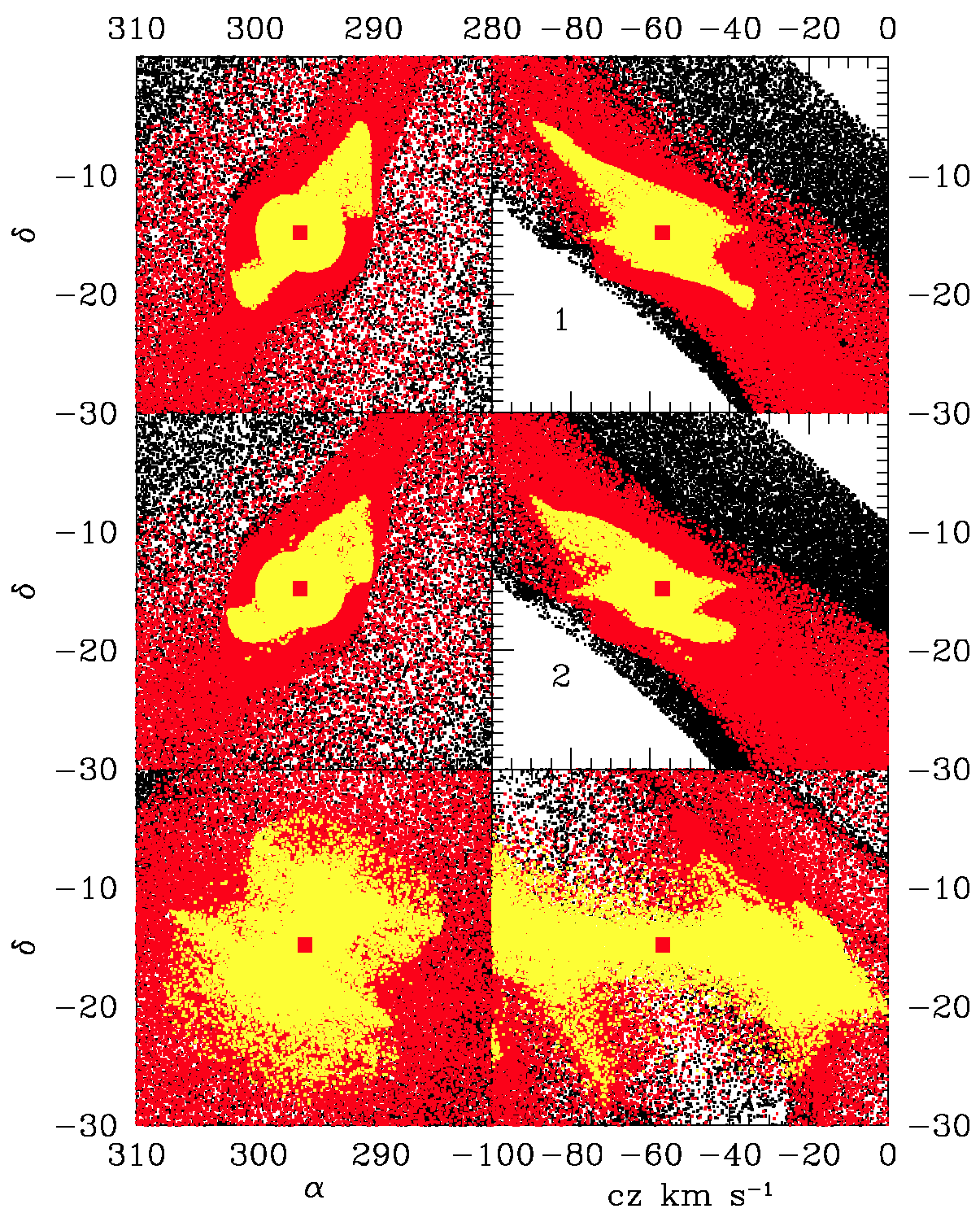} 
\caption{\label{Fig:4} \footnotesize  Model streams around NGC\,6822 in equatorial coordinates for the three solutions in the color scheme of Figure~\ref{Fig:2}. The nominal redshift $cz$ is the component of the heliocentric velocity in the direction to NGC\,6822. Solutions~1 and~2 are labeled; 3 is in the bottom panels.}
\end{center}
\end{figure}

\subsection{A Stream around NGC\,6822}\label{sec:3.2}

The atomic hydrogen around the LG galaxy NGC\,6822 extends to angular radius $\sim 30^\prime$, projected separation $r\sim 4$~kpc. The H\,I redshifts of the  outer parts differ from the center by $v\sim \pm 50$ km~s$^{-1}$. These quantities define a characteristic mass, $v^2r/G\sim 3\times 10^9M_\odot$. The catalog mass of this galaxy, $6\times 10^9M_\odot$, is not well tested by the dynamical model because it is too small to have much effect on the other actors, but its similarity to $v^2r/G$ does suggest that the H\,I could be gravitationally bound to the galaxy. And, if the H\,I were not bound, the relative velocity of $50$ km~s$^{-1}$ would soon carry the hydrogen far beyond its projected separation from the galaxy, unless the relative motion of galaxy and gas were directed almost exactly along the line of sight, which seems unlikely. Thus we ought to study the formation of a gravitationally bound H\,I cloud. This cannot be simulated by the simple gravitational motions of test particles, but we can consider initial conditions for dissipative contraction that might produce the H\,I envelope of NGC\,6822. 

Figure~\ref{Fig:4} shows angular distributions and radial velocities of the primeval streams of test particles initially uniformly distributed around NGC\,6822 in the three solutions. The particles initially less than $r_g/2 = 17$~kpc from NGC\,6822 are plotted in yellow, those initially at 17~to 25~kpc in red, and those at~25 to 34~kpc in black. Here again the cutoff at $r_g$ in the initial  test particle distribution produces the sharp edges in the figure. The range of angles plotted in Figure~\ref{Fig:4} is large enough that the heliocentric motion of the galaxy causes a significant variation of the line-of-sight velocity across the figure. This effect is removed by plotting an effective redshift $cz$  defined as the component of the heliocentric velocity of each test particle along the heliocentric direction to the center of NGC\,6822. 

 The orbital histories of NGC\,6822 in Solutions~1 and~2 are similar (Fig.~\ref{Fig:1}), as are the present test particle distributions in Figure~\ref{Fig:4}, though there are systematic differences. The stream position angles in these two solutions are about $140^\circ$ (measured from the direction of increasing declination toward the direction of increasing right ascension), similar to the orientation of the H\,I stream around this galaxy (Roberts 1972; de Blok \& Walter 2000). In these two model streams the redshift increases with decreasing declination and increasing right ascension, in the direction of the observations, though the gradient is much smaller than observed. This is in line with the idea that the primeval stream may model the precursor to dissipative  contraction. Solution~3 has a different history (Fig.~\ref{Fig:1}) and different present distributions of positions and redshifts that seem less promising. 

\begin{figure}[htpb]
\begin{center}
\includegraphics[angle=0,width=3.5in]{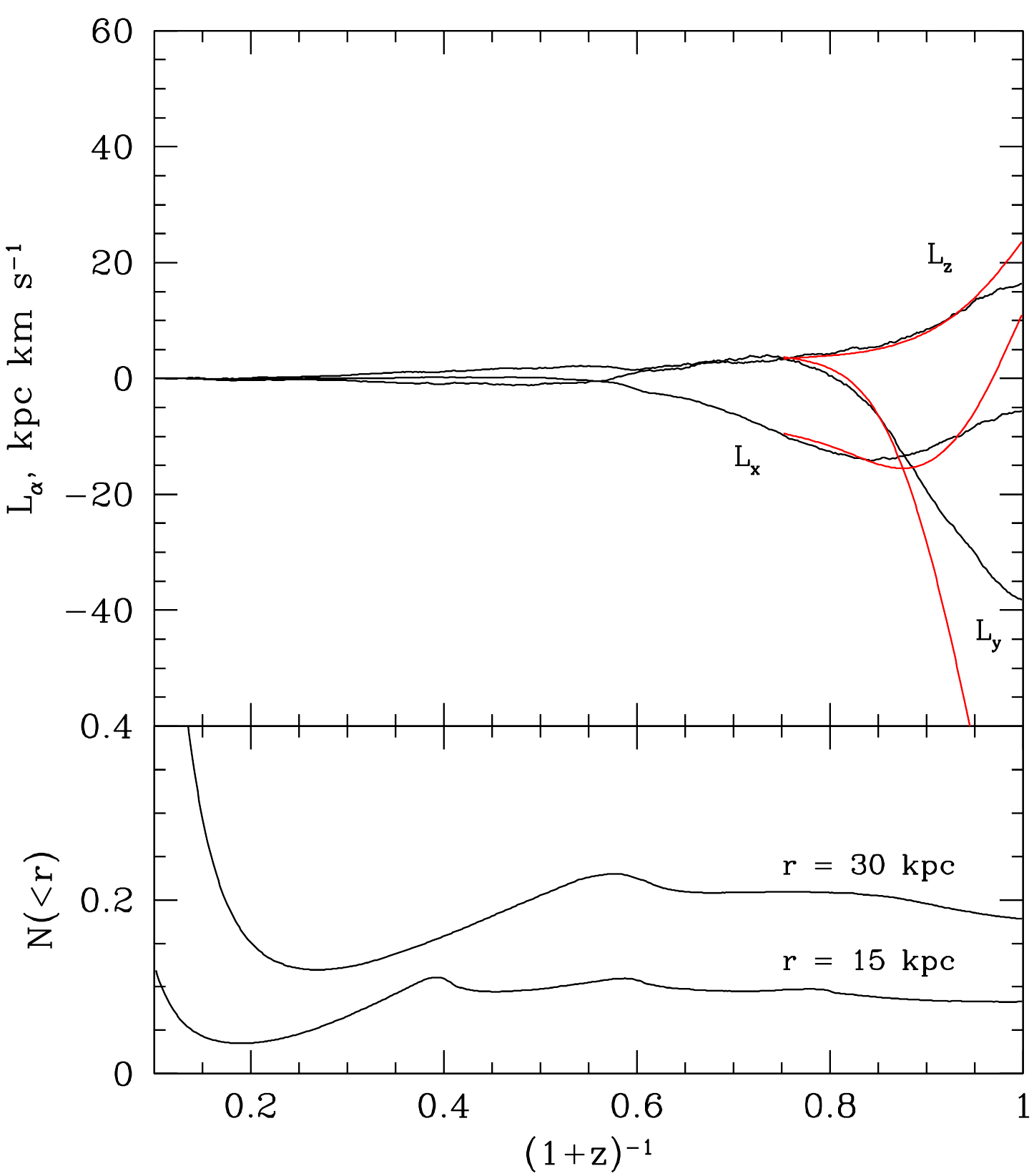} 
\caption{\label{Fig:5} \footnotesize  Evolution of the cloud of test particles around NGC\,6822 in Solution~1. The lower panel shows test particle concentrations.  Black curves in the upper panel are components of mean angular momentum of the particles within 30~kpc. Red curves are components for the fixed set of particles that are within 30~kpc at redshift $z=1/3$.}
\end{center}
\end{figure}

The lower panel in Figure \ref{Fig:5} shows the evolution in Solution~1 of the numbers of particles within physical distances 15 and 30~kpc from NGC\,6822,  plotted as the ratio to the number initially within 30~kpc. The concentrations initially decrease because the cloud is expanding with the general expansion of the universe. The first minima, at $z\sim 4$ for $r=15$~kpc and $z\sim 3$ for  $r=30$~kpc, are artificially deep because the initial velocities relative to NGC\,6822 are artificially radial. At $z<1$ the concentration within 15~kpc is nearly constant at about the value at $z=9$, and the concentration within 30~kpc is about one fifth its initial value.

The black curves in the upper panel of Figure \ref{Fig:5} show the evolution of the components (in galactic coordinates) of the mean (specific) angular momentum per particle relative to the position and motion of NGC\,6822 for the particles that are at physical distance $r<30$~kpc from NGC\,6822. The mean angular momentum at $r<15$~kpc is smaller but the components evolve in a similar way. The angular momentum evolves in part because particles are streaming past the  30~kpc limiting distance, and in part because of the torques from other actors. These effects are separated by identifying the particles that at $(1+z)^{-1}=0.75$ are within 30~kpc from NGC\,6822. The components of mean angular momentum of this fixed set of particles are plotted at lower redshift as the red curves in Figure~\ref{Fig:5}. The red and black curves differ because of the motions of particles through $r=30$~kpc. The similarities show that gravitational torques substantially affect the mean angular momentum per particle near NGC\,6822 as it lingers near its maximum distance from MW and M\,31 approaches. The angular momentum may be compared to the maximum to be expected from the tidal torque by MW integrated over a Hubble time, $L\sim GMr^2/(H_oR^3) \sim 500$ kpc~km~s$^{-1}$ for MW mass $M\sim 10^{12}M_\odot$, NGC\,6822 distance $R\sim 500$~kpc, and  moment arm $r\sim 30$~kpc. This is an order of magnitude larger than the mean angular momentum within $r=30$~kpc at $z=0$ in the model, $L\sim 50$ kpc~km~s$^{-1}$.

At $z=0$ and $r=30$~kpc the angular momentum vector has position angle $\sim 220^\circ$ and inclination $i\sim 55^\circ$ (where $i$ is the angle between the angular momentum vector and the direction from NGC\,6822  to MW, meaning the angular momentum is tilted from the plane of the sky toward us by about $35^\circ$). Over the range of limiting radius $r=10$~kpc to $100$~kpc the direction of the model angular momentum vector does not change much, but the magnitude increases with increasing $r$ from about 20 kpc~km~s$^{-1}$ at $r=10$~kpc to about 200 kpc~km~s$^{-1}$ at $r=100$~kpc. The observable component of angular momentum per unit mass in the H\,I around NGC\,6822 is about 100 kpc~km~s$^{-1}$. The position angle is similar to the  model stream, perhaps a significant coincidence. It will be noted, however, that the model stream acquired its angular momentum at redshift $z\la 0.3$, so it could be a progenitor of the H\,I cloud around NGC\,6822 only if the accretion of this H\,I were a recent development. This special condition is discussed in Section~\ref{sec:4}. 

\begin{figure}[htpb]
\begin{center}
\includegraphics[angle=0,width=2.75in]{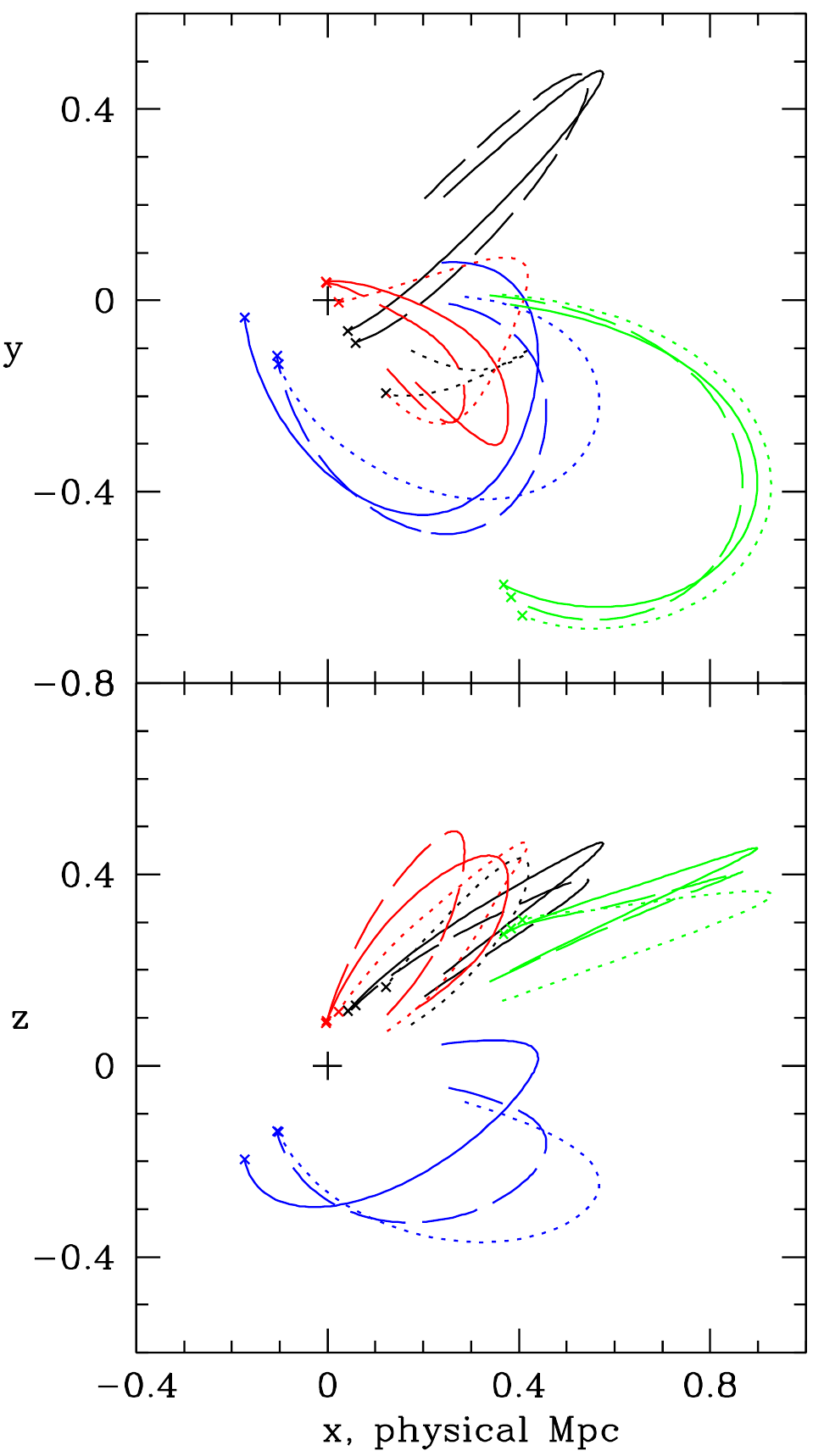} 
\caption{\label{Fig:6} \footnotesize   Orbits around M\,31, at the origin, for M\,33 (blue), NGC\,185 (black), NGC\,147 (red), and MW (green). Solution 1 is plotted as the solid line, 2 as long dashes, and 3 as dots.}
\end{center}
\end{figure}

\subsection{Streams around M\,31}\label{sec:3.3}

The streams around M\,31 are dominated by stars (Richardson, Irwin, McConnachie, et al.\ 2011, and references therein), a different situation  from the loose stream of H\,I around the Magellanic Clouds or the gravitationally bound H\,I around NGC\,6822. The prominent optical stream between M\,31 and M\,33 invites the idea that these two galaxies suffered a close passage, which did not happen in the LG solutions used here. This is illustrated in Figure~\ref{Fig:6}, which shows in galactic coordinates the paths of galaxies relative to the position of M\,31 at the plus sign. The green curves show MW approaching M\,31 from the right, in the mirror image of the approach of M\,31 to MW in Figure~\ref{Fig:1}. The orbits of NGC\,185 (black) and NGC\,147 (red) are similar in Solutions~1 and~2 and rather different in Solution (3). The situation may be compared to NGC\,6822 in Figure~\ref{Fig:1}. In all three solutions M\,33 (blue) has been well away from M\,31 and MW. But we can offer an illustration of how M\,33 and M\,31 might be connected by a primeval stream. 

\begin{figure}[htpb]
\begin{center}
\includegraphics[angle=0,width=4.75in]{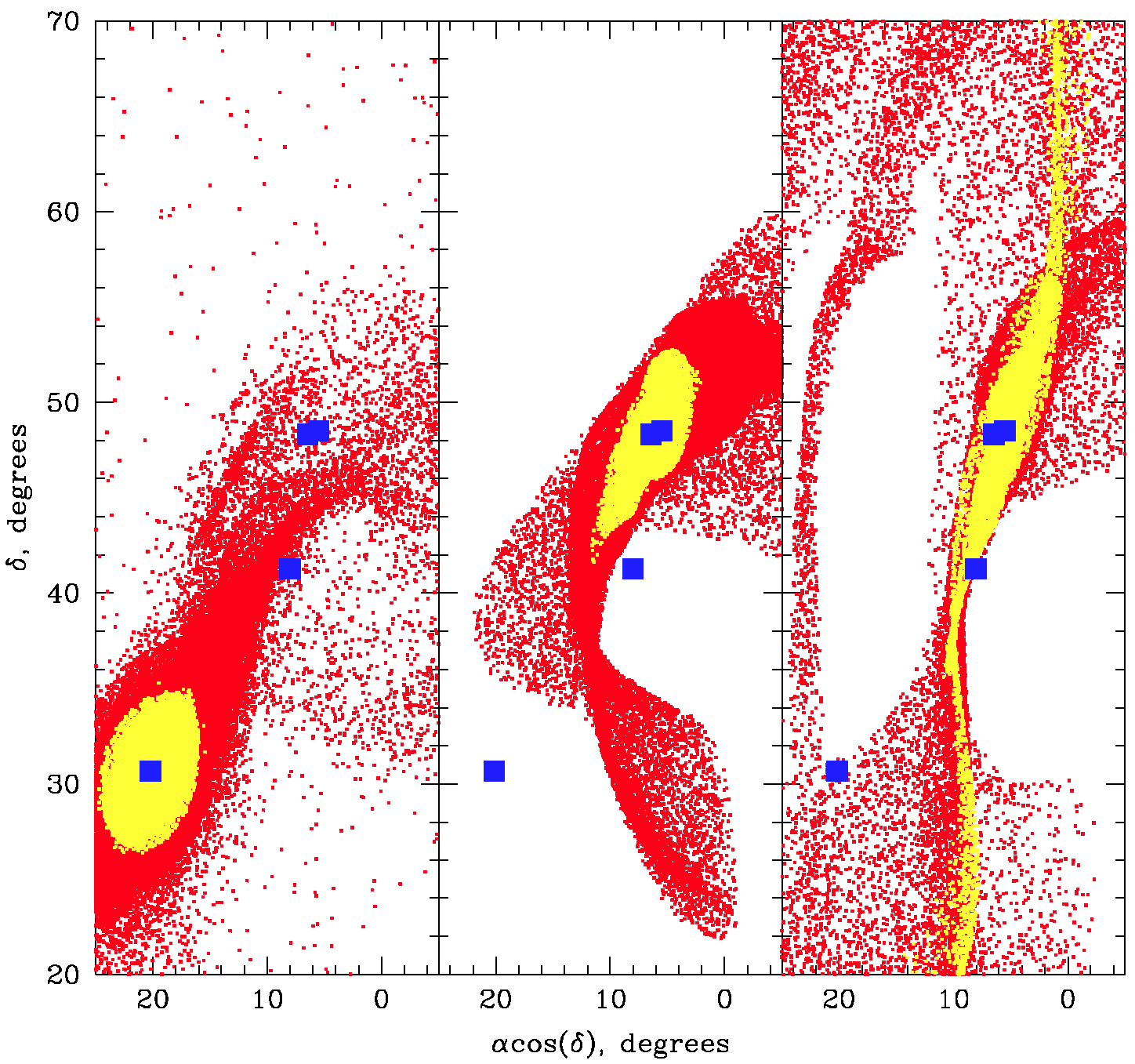} 
\caption{\label{Fig:7} \footnotesize  Present distributions of the test particles in Solution 1 that at $z=9$ were close to M\,33, now at the blue square to the lower left in the left panel, initially close to NGC\,185, now at the left of the top two squares in the middle panel, and NGC\,147, at the right uppermost square in the right-hand panel. M\,31 is at the center square.} \end{center}
\end{figure}

Figure~\ref{Fig:7} shows streams of test particles around M\,33 in the left panel, NGC\,185 in the central panel, and   NGC\,147 in the right-hand panel. The test particles plotted as yellow were initially closer to the galaxy than $r_g/2$, and red initially at $r_g/2$ to $3r_g/4$, where $r_g \sim 100$~kpc in M\,33 and $\sim 40$~kpc in NGC\,185 and NGC\,147. Test particles initially further out add more diffuse streaks. To reduce clutter we refrain from including IC\,10 and the stars that might have been pulled out of M\,31. If stars formed around the young M\,33 and were drawn away by tidal fields of MW, which was 230 kpc away at $z=9$, and M\,31, which was 250~kpc from M\,33, then the left-hand panel in Figure~\ref{Fig:7} suggests the stars might end up in an observable stream that passes across M\,31, at the central square in the figure. A primeval stream of stars drawn from the outskirts of NGC\,185 could appear somewhat tighter, and a stream from NGC\,147 could be tighter still, though both avoid M\,33. 

\section{Discussion}\label{sec:4}

The plausible approximation to MS in Figure~\ref{Fig:2} is based on a dynamical model that fits the considerable number of measurements in Tables~1 and~2 about as well as could be expected. This dynamical model offers a reasonably unambiguous prediction of the positions of the galaxies near LMC at high redshift. The primeval stream model assumes  the young LMC had an envelope of H\,I or cool plasma whose  response to the presence of the neighboring galaxies at high redshift may be approximated by a cloud of massless test particles initially moving with LMC. It also assumes memory of the conditions at high redshift is preserved in the present state of the H\,I. Memory in a cloud of test particles is illustrated in Figure~\ref{Fig:3}. Besla et al. (2010) summarize arguments that MS is ``a young feature (1 - 2~Gyr)''. In the primeval stream model LMC would have entered an MW corona that extends to 300 kpc at redshift $z\sim 0.1$, or about 1~Gyr ago, which may be  recent enough that the trailing component largely survived moving through the plasma while losing much of the leading component. Prior to that LMC would have been well separated from large galaxies and its proto-MS might have survived in the same manner as the H\,I at similar surface densities around other isolated gas-rich dwarfs. These are not many assumptions, and they are applied in a straightforward way, which lends support for the result. A test by hydrodynamic simulation could provide a stronger argument, but the evidence we have now is that gravitational interactions among the young galaxies can have produced MS. 

We must consider that plausible MS models have been obtained  without reference to initial conditions, and from a variety of ways to model the orbits and interactions among LMC, SMC and MW (Gardiner \& Noguchi 1996; Mastropietro  et al. 2005; Connors, Kawata, \& Gibson 2006; Diaz \& Bekki 2011, 2012; Besla, Kallivayalil, Hernquist, et al. 2010, 2012). Though these approaches generally require close attention to parameter choices, it appears that MS has properties of an attractor, capable at arriving at a good approximation to its present state --- and, we must expect, of being destroyed --- under a variety of interactions along the way. The case that MS originated as a primeval stream rests on the demonstration that such a stream would have formed if cool baryons in the young galaxy were in a position to be tidally disturbed and then not seriously disturbed thereafter. We must be cautious, however, for MS seems to have a generic tendency to end up looking like Figure~\ref{Fig:2}. 

In contrast to MS, the case for formation of the H\,I envelope of NGC\,6822 out of a primeval stream requires a very special situation. However, the situation may be indicated by the demonstration by Demers, Battinelli, \& Kunkel (2006) that NGC\,6822 is a polar ring galaxy. The stellar halo traced by RGB stars extends about as far from the galaxy as the H\,I envelope, but the long axis of the stellar distribution has position angle $60^\circ$, while the H\,I has position angle is $130^\circ$ (de Blok \& Walter 2000). The redshift gradients of the halo stars and the H\,I envelope both point along the long axis of their angular distribution, and both gradients are close to constant at 15~km~s$^{-1}$~kpc$^{-1}$. The contributions to the angular momenta of stars and H\,I by the observed redshift gradients have position angles differing by $\sim 70^\circ$. This could signify a strong departure from axial symmetry, as in bars. (Hodge 1977 notes that the stars in the inner $\sim 0.5$~kpc, with PA$\sim 10^\circ$, may be a bar.) The alternative is that the angular momenta of stellar halo and H\,I envelope have quite different directions. Similar tilts of the long axis of the stellar distribution from the H\,I redshift gradient  are observed in isolated dwarfs (Stanonik, Platen, Arag{\'o}n-Calvo, van Gorkom, et al. 2009;  Kreckel, Peebles, van Gorkom, van de Weygaert, \& van der Hulst 2011).  This might indicate that the H\,I envelopes dissipatively settled onto these galaxies without adding many stars. 

The orientation and the direction of the redshift gradient  in the primeval stream around NGC\,6822 in Solutions~1 and~2 (Fig.~[\ref{Fig:4}]) agree with the H\,I envelope of this galaxy. The model redshift gradient is much smaller than observed, but that could be because of the dissipative contraction of the H\,I. This picture requires that the H\,I settled after redshift $z\sim 0.3$, when the primeval stream acquired its angular momentum (Fig.~[\ref{Fig:5}]), a very special condition that may be crudely modeled as follows. Suppose that at $z\sim 0.3$ the stars in NGC\,6822 were centered on a sheet of diffuse baryons with number density $n_b\sim 10^{-4}$~cm$^{-3}$ and thickness $h\sim 30$~kpc, or angular width $\sim 3^\circ$ at the distance of this galaxy, which is on the scale of what is plotted in Figure~\ref{Fig:4}. The characteristic baryon surface density is  $\sim n_bh\sim 10^{19}$~cm$^{-2}$, or baryon surface mass density $\Sigma_b\sim 10^5M_\odot$~kpc$^{-2}$. With total surface density $\Sigma_m\sim6\Sigma_b$, to take account of dark matter, pressure support requires plasma temperature $kT\sim 2\pi G\Sigma_mm_ph\sim 3$~eV. That makes the plasma cooling time ($\tau\sim 10^{11.4} \sqrt{T}/n_b$ in cgs units, ignoring line emission) about $10^{10}$~y, roughly what is wanted for late accretion of an H\,I envelope around NGC\,6822. If an $h$ by $h$ (30~kpc by 30~kpc) piece of the slab collapsed by a factor of three in each direction it would gather baryon mass $\Sigma_bh^2\sim 10^8M_\odot$ in a region of width $\sim 10$~kpc $\sim 1^\circ$ at NGC\,6822, roughly what is detected in 21-cm radiation (de Blok \& Walter 2000). Conservation of angular momentum would bring the redshift gradient in Figure~\ref{Fig:4}, 0.4~km~s$^{-1}$~kpc$^{-1}$, up by a factor of 9 to $\sim4$~km~s$^{-1}$~kpc$^{-1}$, approaching what de Blok and Walter observe. The point of this crude set of estimates is that late collapse might happen around NGC\,6822 as well as  other polar ring galaxies. 

Our numerical method of finding the orbits of LG actors is not efficient at arriving at solutions in which actors have orbited each other several times. This is not a problem for the motion of LMC around MW, because the present conditions seem to be well enough known to exclude multiple orbit passages (Besla, Kallivayalil, Hernquist, et al.\ 2007). It does not seem to be a problem for NGC\,6822, either, because its present slow motion away from MW with standard estimates of the MW mass make it likely that NGC\,6822 has not completed more than one orbit. The limitation of the numerical method is more serious for the smaller galaxies now near M\,31. In the Local Group solutions used here the pair of dwarf spheroidal galaxies NGC\,185 and NGC\,147 has mass $\sim 10^{10}M_\odot$, above van den Bergh's (1998) estimate of what is required if these galaxies are a bound system, and indeed our solutions show the two galaxies completing about one orbit of relative motion. A solution with completion of several orbits, perhaps leaving trails of stars, could have been missed, however. The dynamical solutions show M\,31 and M\,33 approaching each other for the first time after separating at high redshift (Fig.~[\ref{Fig:6}]). That does not agree with the idea that the optical stream between M\,31 and M\,33 is a remnant of a close passage of the two galaxies.  Perhaps we have not found the right orbit for M\,33. Perhaps the stream between M\,33 and M\,31  formed by close passage of one of the other galaxies near M\,31, as discussed by Sadoun, Mohayaee, \& Colin (2013). Or perhaps this is a primeval stream that happened to have been loaded with stars. We must add that the primeval stream picture requires the special postulate that stars are pulled out of the young galaxies in and around M\,31, while H\,I would have been pulled out the Magellanic Clouds to make the Magellanic Stream. But Figure~\ref{Fig:7} does show that primeval streams can run across M\,31, which may merit closer consideration.

\acknowledgments
We have benefited from discussions with Ed Shaya and support from the NASA Astrophysics Data Analysis Program award NNX12AE70G and from a series of awards from the Space Telescope Science Institute, most recently associated with programs AR-11285, GO-11584, and GO-12546.

\end{document}